\documentclass[sigconf,nonacm,balance=false,pbalance=false]{acmart}
\usepackage{booktabs,array,graphicx,amsmath,algorithm,algpseudocode,xcolor}
\usepackage{flushend}
\usepackage{placeins}
\newcommand{\sys}{LADS}

\newcommand{\resultblank}{\textcolor{red}{\underline{\hspace{0.6cm}}}}
\newcommand{\metric}[1]{\resultblank}

\hypersetup{pdfauthor={Yuxuan Hou, Linshan Jiang},pdftitle={LADS: Lease-Aware Delivery Scheduling of Tool Results for Language Model Agents}}
\begin{document}
\title{LADS: Lease-Aware Delivery Scheduling of Tool Results for Language Model Agents}
\author{Yuxuan Hou}
\affiliation{%
  \department{Department of Computer Science\\and Engineering}
  \institution{Southern University of Science and Technology}
  \city{Shenzhen}
  \country{China}
}
\author{Linshan Jiang}
\affiliation{%
  \department{Department of Computer Science\\and Engineering}
  \institution{Southern University of Science and Technology}
  \city{Shenzhen}
  \country{China}
}
\renewcommand{\shortauthors}{Yuxuan Hou and Linshan Jiang}
\date{}
\begin{abstract}
Web agents complete multi-step online tasks through repeated browser actions and model calls. Their returned observations lie on the critical path to subsequent decisions. When concurrent browser actions finish in bursts, observations can accumulate on a constrained return path to a shared inference service, while waiting histories compete for reusable cache state. Conventional delivery rules prioritize short transfers or immediate cache reuse but overlook how sending a result earlier can help a request resume sooner and free cache-protection capacity for other waiting requests. We present \sys{} (Lease-Aware Delivery Scheduling), a tool-result scheduler that couples cache-aware delivery with bounded prefix leases. Its key insight is that resumption lets extra protection references retire, making lease quota available to protect another waiting history. \sys{} prioritizes timely handoffs using current lease ownership and an estimated resumption delay, with bounded departures from Cache-first order. Compared with ReAct using first-in, first-out (FIFO) tool-result delivery, \sys{} reduces mean complete-workflow latency by 44.6\% under bursty tool returns and by 42.4\% on Web-derived burst replay.
\end{abstract}
\maketitle
\hypersetup{pdfauthor={Yuxuan Hou, Linshan Jiang}}
\begingroup
\section{Introduction}
\label{sec:intro}

Web agents advance through observations: a returned page, retrieved passage, or tool response supplies the context for the next model decision. ReAct makes this reasoning--action--observation loop explicit~\cite{yao2023react}. When concurrent agents share an inference service, completed observations also share a return path. Waiting on that path delays the next model call and every dependent action. We study how to order these completed results to reduce complete-workflow latency.

The difficulty is that delivery order affects both communication waiting and the computation needed to resume an agent. Concurrent tool completions can create a backlog on a constrained return path, while their model histories compete for reusable key--value (KV) cache state. First-in, first-out (FIFO) follows readiness order; Size-first advances short transfers; Cache-first prioritizes results whose associated history prefixes are currently resident in the KV cache. A history observed in cache must still survive committed traffic and engine waiting before reuse. The useful scheduling signal therefore extends beyond result size or an immediate cache hit.

Existing systems offer controls at several points in this process. PagedAttention supplies block-based KV management~\cite{kwon2023pagedattention}; InferCept and Continuum manage state across interruptions and multi-turn execution~\cite{abhyankar2024infercept,li2026continuum}. Application-aware serving uses dependencies and program progress to schedule model calls~\cite{lin2024parrot,luo2026agentix}. Tail-aware Release separates readiness from commitment and combines priority with feedback-driven control of committed model work~\cite{feng2026release}. Before completed tool results are committed for transmission, a further question arises: how can delivery order preserve reusable histories and make bounded prefix-cache protection available to other waiting agents?

Our key insight is that engine-native resumption---handled by vLLM's own scheduler and block manager---can return protection quota before inference finishes. A bounded prefix lease adds references to keep a resident history available while its observation waits. Once the inference engine acquires that history and allocates new tokens, the extra references can retire. Their logical quota can protect another waiting history while the resumed request continues using its physical cache blocks. Delivery can thus influence who receives protection next. Exploiting this opportunity requires identifying owners with exclusive quota and estimating whether their results can reach allocation before the leases expire.

We present \sys{} (Lease-Aware Delivery Scheduling), which connects result ordering to bounded prefix protection. Its cache-aware base order favors currently reusable histories. Leases are granted in FIFO order with a fixed maximum lifetime.

The protection budget is based on expected near-term prefix reuse and capped to reserve cache space for model execution. A turnover priority advances a protected result when an estimated timely handoff can return exclusive quota for another waiting history. Each actual promotion is followed by a base-order choice, bounding consecutive interventions. Fully received observations continue through ordinary FIFO model entry; the engine verifies native ownership and positive-token allocation before retiring the extra references.

\begin{figure*}[t]
\centering
\includegraphics[width=.92\textwidth,trim=4.9bp 5.2bp 4.9bp 2.7bp,clip]{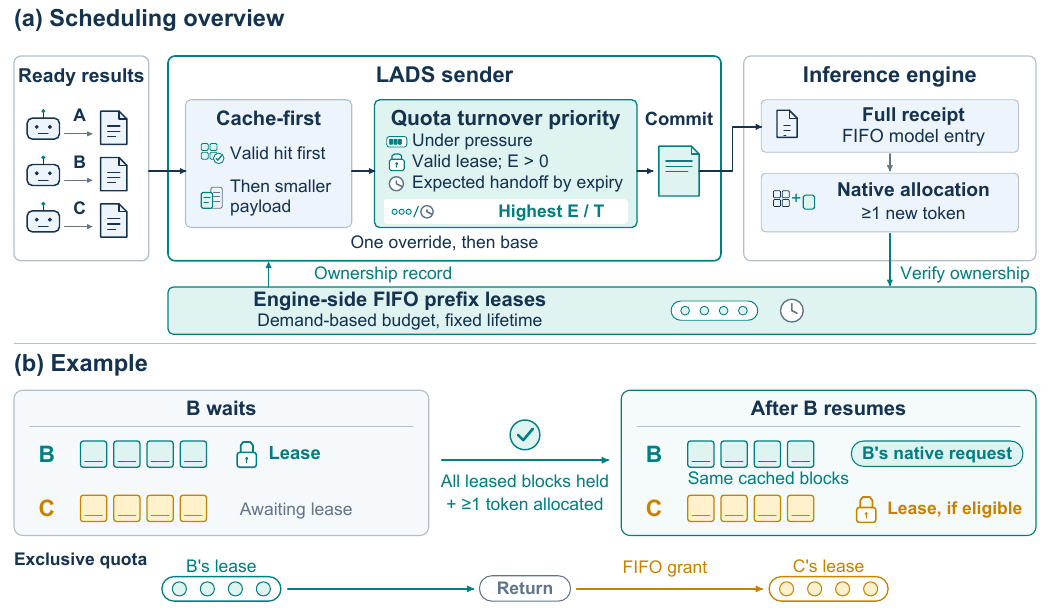}
\caption{LADS scheduling and quota turnover. A, B, and C label the tool results and associated model histories of three concurrent Web agents. (a) Cache-first chooses among results permitted by the transport window. Under pressure, eligible owners are ranked by $E/T$, where $E$ is exclusive protection quota and $T$ is estimated handoff time. Each override is followed by a base choice; committed frames retain their order. FIFO leases use a demand-based budget and fixed lifetime limit. (b) Two cache states show the same cached blocks. After B's native request owns every leased block and successfully allocates at least one new token, its extra lease references retire. B keeps using its blocks. Returned exclusive quota can fund a FIFO lease on C's own resident prefix if grant conditions hold.}
\label{fig:overview}
\Description{A, B, and C identify three concurrent Web agents' tool results and associated model histories. Panel a retains the current scheduling overview: Cache-first selects by default, eligible lease owners can receive bounded turnover priority under protection pressure, and committed results enter ordinary FIFO registration and allocation. Engine-side FIFO leases use a demand-derived budget and fixed lifetime. Panel b shows two states of the same resident cache. The B and C rows retain identical block counts, positions and colors across time. Initially B has a lease and C awaits one. After verified handoff, B's native request owns every leased block and has successfully allocated at least one new token, allowing the extra lease references to retire while the same physical blocks remain in use. A separate lower rail uses circles for exclusive logical protection quota, showing its return from B's lease and possible FIFO grant to C's own resident prefix. The rectangular blocks are never transferred from B to C. C receives a lease only if all grant conditions hold.}
\end{figure*}

We evaluate the complete method and its components through controlled bursts and Web-derived burst replay, and additionally validate delivery on captured Web-agent task paths. The two main workloads compare LADS with FIFO, Size-first, and a delivery-side adaptation of Tail-aware Release retaining its work accounting and feedback. Web-derived replay supplies Web observations within the controlled workload construction. Across five input seeds in the controlled cohort-8 realization, \sys{} reduces mean complete-workflow latency by 44.6\% relative to FIFO. On Web-derived burst replay, the corresponding reduction is 42.4\% across three input seeds.

\begingroup
Our main contributions are summarized as follows:
\begin{itemize}
\item We introduce \sys{}, a tool-result delivery scheduler that reduces complete-workflow latency for concurrent Web agents by coordinating result ordering with prefix-cache protection.
\item We identify resumption as an opportunity to return cache-protection quota before inference completes. Bounded prefix leases and an allocation-time ownership check make this quota available to other waiting histories while active requests continue using their cached prefixes.
\item Across five input seeds in the controlled cohort-8 realization and three input seeds in Web-derived burst replay, \sys{} reduces mean complete-workflow latency by 44.6\% and 42.4\%, respectively, relative to first-in, first-out delivery.
\end{itemize}
\endgroup

\par
\endgroup
\section{Background and Motivation}
\label{sec:background}

\subsection{The tool-result return path}
\label{sec:return-path}

Concurrent Web agents return completed observations through a shared path to a shared inference engine. The sender chooses among ready, uncommitted result frames; committed frames retain their transport order. Complete receipt immediately registers the continuation in FIFO order, leaving batching and execution to the engine. Delivery order affects both resumption delay and whether cached histories survive until reuse.

\begin{figure}[t]
\centering
\includegraphics[width=.95\columnwidth,trim=2.1bp 6.3bp 0.9bp 2.9bp,clip]{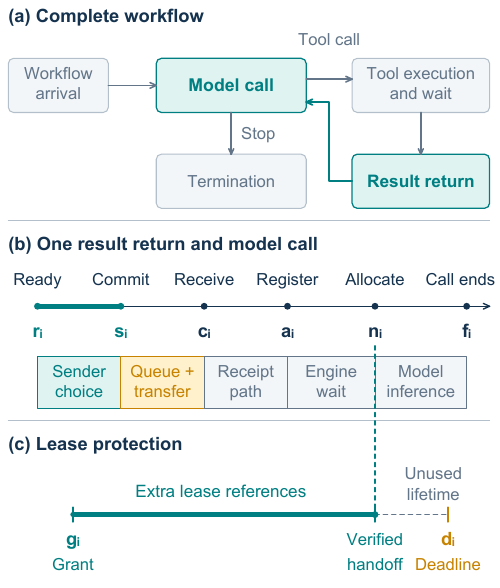}
\caption{Schematic workflow, return path, and prefix protection. (a) Model calls alternate with tool use until termination. (b) LADS selects a ready frame before commitment at $s_i$; committed queuing and transfer end at $c_i$. (c) Lease references run from $g_i$ to verified handoff at $n_i$ or an earlier lease-termination event. The dashed segment is unused lifetime before $d_i=g_i+\ell$. Grant and deadline need not coincide with readiness and model-call completion.}
\label{fig:return-path}
\Description{Three corresponding panels. Panel a shows workflow arrival, a model call, a tool-execution-and-wait branch, result return, and a loop back to the model; a separate stop branch leads to workflow termination. The teal result-return and following model-call nodes in panel a are detailed in panel b. Its events are result readiness r, full-frame commitment s, complete receipt c, request registration a, first successful positive-token allocation n, and the end of that model call f. The pre-commit interval marks sender choice; subsequent intervals are committed queuing and transfer, receipt processing, engine waiting, and model inference. Panel c uses the same horizontal time coordinates. Grant g is distinct from r, and the cumulative deadline d equals g plus lifetime ell is distinct from f. Extra lease references run from g to verified handoff. A vertical guide aligns handoff with the allocation callback at n; within that callback, successful allocation precedes native ownership verification and release of extra references. This is a successful handoff illustration, not an assertion of identical internal timestamps. The dashed segment from handoff to the deadline is unused lease lifetime. Earlier termination events can end protection sooner. All timing is schematic.}
\end{figure}

Each completed tool result is encapsulated in an application-layer result frame containing the result payload and its metadata. Let $r_i$, $s_i$, $c_i$, $a_i$, $n_i$, and $f_i$ denote result readiness, full-frame commitment, complete receipt, request registration, first successful positive-token allocation, and inference completion for continuation $i$. Sender waiting spans $[r_i,s_i]$, while $[c_i,n_i]$ includes receipt processing and engine waiting. These intervals jointly determine how long a cached history must survive before reuse. Complete-workflow latency is measured from workflow arrival to termination and includes all model calls and tool waits. Figure~\ref{fig:return-path} locates the delivery decision and the prefix-survival interval.

\subsection{Prefix leases and the turnover opportunity}

Block-based KV management and multi-turn context caching permit the service to reuse resident history instead of recomputing it~\cite{kwon2023pagedattention,gao2024cachedattention}. A prefix lease temporarily adds references to already-resident blocks while their continuation waits. It preserves available state rather than reconstructing evicted state.

\paragraph{Motivating example.} Suppose ready results A and B hold the available leases, while C's resident history awaits protection. Cache-first selects A's smaller frame. If B owns enough exclusive quota to protect C, returning B first can make that quota available earlier than sending B after A, provided B reaches allocation before its lease expires. C can then obtain a lease while B continues inference on the same physical blocks.

For lease $i$, let $L_i(t)$ be the set of KV blocks it protects at time $t$, and $d_i$ its expiry time. Let the protection occupancy $U(t)$ count the distinct blocks protected across all leases. The admission target $B(t)$ governs new lease grants, while the protection commitment $C(t)$ accounts for both this target and existing protection. Let $K$ be the usable block-pool capacity and $H$ the headroom reserved for model allocation. All budgets and block counts are measured in KV blocks:
\begin{equation}
 \begin{aligned}
 U(t)&=\left|\bigcup_i L_i(t)\right|,\\
 C(t)&=\max\{B(t),U(t)\},\quad U(t)\leq C(t)\leq K-H.
 \end{aligned}
 \label{eq:lads-budget}
\end{equation}
Shared blocks count once toward $U(t)$. Lowering $B(t)$ does not itself revoke existing leases; when $U(t)>B(t)$, $C(t)=U(t)$ accounts for the protection still outstanding. Let $E_i(t)$ count the blocks protected by lease $i$ and by no other lease:
\begin{equation}
 E_i(t)=\left|\left\{b\in L_i(t):\sum_j\mathbf{1}[b\in L_j(t)]=1\right\}\right|.
 \label{eq:lads-exclusive}
\end{equation}
Here $\mathbf{1}[\cdot]$ equals 1 when its condition holds and 0 otherwise. The sum over $j$ counts the leases protecting block $b$; requiring it to equal 1 selects blocks protected only by lease $i$.

When the engine acquires its own prefix references and successfully allocates positive-token work, the extra lease references can retire. This handoff can return $E_i$ units of \emph{logical protection quota}, allowing another resident prefix to obtain a lease. The model may still use the same physical blocks. Quota turnover is therefore distinct from freeing GPU memory or completing inference.

\section{LADS Design}
\label{sec:design}

\subsection{Problem definition}

We consider concurrent Web agents that share an inference service and a return path for completed tool observations. Bursty returns can create a transmission backlog while waiting histories compete for cache capacity.

At time $t$, let $\mathcal{F}(t)$ contain ready, uncommitted result frames permitted by the transport window. Given payload sizes, cache hints, and published lease state, the sender selects the next frame for commitment to reduce complete-workflow latency under bounded prefix protection.

\subsection{Design overview}

LADS combines a demand-based admission target, FIFO prefix leases, and bounded departures from Cache-first order (Figure~\ref{fig:overview}). The engine publishes lease ownership, deadlines, and exclusive quota. Under protection pressure, the sender advances an eligible owner whose estimated handoff precedes expiry; verified handoff makes that quota available to FIFO replenishment.

\subsection{Cache-first default ordering}

For each $i$, the sender knows payload size $p_i$, a version-bound cached-prefix hint $\widetilde C_i$, and a stable identity $\mathrm{id}_i$. Its default selection minimizes
\begin{equation}
 k_i=\bigl(\mathbf{1}[\widetilde C_i=0],\ p_i,\ \mathrm{id}_i\bigr),
 \label{eq:lads-base}
\end{equation}
lexicographically, treating invalid hints as zero. The key prioritizes valid nonzero cache hits, then smaller payloads and stable identity. This is the default whenever no usable turnover opportunity exists.

\begin{figure}[t]
\centering
\includegraphics[width=.95\columnwidth,trim=2.1bp 2.9bp 3.9bp 3.9bp,clip]{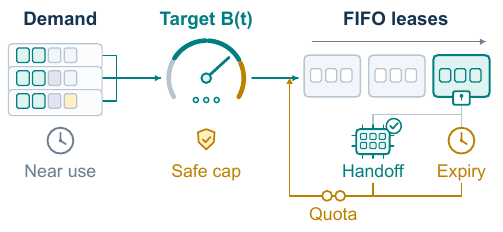}
\caption{Demand-based protection and FIFO admission. Near-use demand and the safe ceiling set the aggregate target $B(t)$, with a whole-prefix fit check. FIFO order selects recipients; each grant protects a complete resident prefix subject to target and headroom checks. Verified handoff retains the cached blocks under native references and returns exclusive logical quota; expiry also ends lease protection. The return loop supports replenishment. A lower target preserves existing leases, with $C(t)=\max\{B(t),U(t)\}$ accounting for outstanding protection.}
\label{fig:budget-admission}
\Description{An open pictorial left-to-right mechanism diagram. Resident-prefix block groups contribute aggregate near-use demand to a segmented target gauge; a horizon clock and a safe-cap shield indicate its time and resource constraints. The target bounds admission from a separate FIFO queue. The rightmost whole-prefix card receives an enclosing lease band and lock; remaining cards wait in ready-time order. Handoff is drawn as an engine retaining cache blocks with a verification mark, and expiry as a clock. A return loop carries round logical-quota tokens back to admission, never physical cache blocks to another history. The distinct-demand, whole-prefix fit, allocation-headroom and target-reduction conditions are explained in the caption and adjacent equations. Eight short labels replace the previous prose-filled cards. All block counts and icon positions are schematic.}
\end{figure}

\subsection{Budget estimation}
\label{sec:lads-budget-estimation}

For each current, noncold pending result $i$, let $P_i(t)$ be its resident reusable-prefix block set. The budget identifies prefixes expected to be reused within the remaining lease horizon and aggregates their distinct blocks. Its admission target $B(t)$ adapts to this demand while reserving headroom for model allocation.

Let $\ell$ be the cumulative lease lifetime limit and $g_i$ the time of result $i$'s first successful lease grant. The remaining horizon is $\tau_i(t)=\ell$ before that grant and $\tau_i(t)=g_i+\ell-t$ thereafter.

The budget estimate follows the current FIFO frame inventory through result $i$. Let $j\preceq i$ denote the frames up to and including $i$, and let $\nu_i$ count its preceding frames. Let $Q_B$ be the known committed-frame backlog in bytes, $h$ the framing bytes per result, $p_j$ its payload bytes, and $R$ the current link rate in bytes/s. Let $w$ and $a$ count engine waiting requests (including preempted requests) and running requests, $m$ be the engine's concurrency limit, and $\kappa$ the frozen wall-time proxy per 512-token chunk. The estimated time to first successful allocation is

\begin{equation}
 \begin{aligned}
 \widehat T_i^B(t)={}&\frac{Q_B+\sum_{j\preceq i}(h+p_j)}{R}\\
 &{}+\left(\left\lfloor\frac{w+a+\nu_i}{m}\right\rfloor+1\right)\kappa.
 \end{aligned}
 \label{eq:lads-budget-delay}
\end{equation}

The first term accounts for committed transport work and the FIFO frame prefix through $i$; the second is a coarse engine-queue delay proxy. This estimate ends at the first successful allocation, before the request's own missing-input computation.

The near-use set $\mathcal N(t)$ contains current, noncold pending results satisfying $P_i(t)\neq\varnothing$, $\tau_i(t)>0$, and $\widehat T_i^B(t)\leq\tau_i(t)$. Its distinct-block demand is $D(t)=|\bigcup_{i\in\mathcal N(t)}P_i(t)|$. With $F(t)$ currently free blocks, the safe ceiling and admission target are

\begin{equation}
 \begin{aligned}
 C_{\mathrm{safe}}(t)&=\max\{0,\min\{K-H,F(t)+U(t)-H\}\},\\
 B(t)&=\begin{cases}
 \min\{D(t),C_{\mathrm{safe}}(t)\},&\text{if a whole prefix fits},\\
 0,&\text{otherwise}.
 \end{cases}
 \end{aligned}
 \label{eq:lads-demand-target}
\end{equation}

The fit condition requires some $i\in\mathcal N(t)$ with $|P_i(t)|\leq C_{\mathrm{safe}}(t)$. Protection starts at zero. The lease provider evaluates the target every 100\,ms and synchronously before a real first offer; periodic evaluation and replenishment share that interval. The near-use set determines aggregate demand; the resulting target is applied to FIFO lease grants. Figure~\ref{fig:budget-admission} summarizes how the target and returned quota support FIFO whole-prefix admission.

\begin{figure*}[t]
\centering
\includegraphics[width=.92\textwidth,trim=6.1bp 3.5bp 5.7bp 5.5bp,clip]{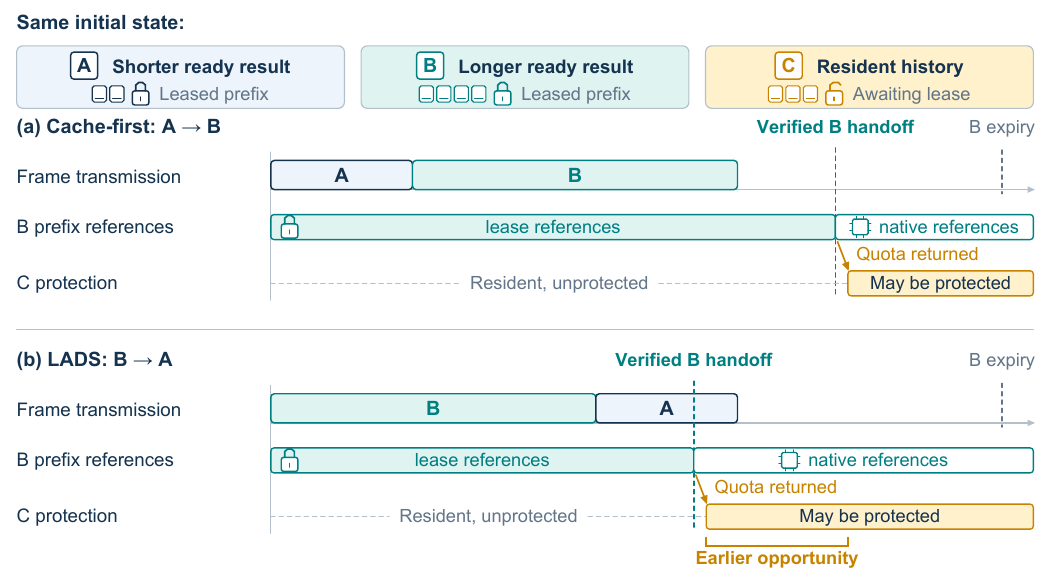}
\caption{Example of earlier delivery creating an earlier protection opportunity. Cache-first sends A's shorter frame; one eligible override sends B first, followed by A. B's verified handoff returns enough exclusive quota to protect C, whereas A's alone cannot. B retains its physical blocks under native references. B's handoff precedes expiry in both orders; C remains resident, is next in FIFO order, and passes target, whole-prefix, and headroom checks.}
\label{fig:turnover-example}
\Description{The two panels share an initial state, a time origin, transmission durations, and the same engine-delay assumptions. A and B hold leases; C's prefix remains resident and awaits protection. Both A and B have valid cache hits, and A's shorter frame makes it the Cache-first choice. Under protection pressure with a valid fresh ownership record, B satisfies lease identity, positive exclusive quota, finite delay, deadline, and highest eligible-score checks. One override sends B before the next base choice A; no intervening arrival changes that base choice, and if C is feasible A ranks ahead of it. The upper panel sends A then B, while the lower sends B then A. B's verified handoff, requiring native ownership of all protected blocks and successful positive-token allocation, happens earlier in the lower panel and before the same B expiry in both panels. B's lease references end at handoff while native references keep its physical blocks in use. A's handoff alone returns insufficient exclusive quota for C, while B's handoff returns enough. A short replenishment interval then grants protection to C's own resident prefix. C is next in FIFO order and passes the recomputed post-handoff target, whole-prefix admission, and actual headroom checks. The diagram shows a conditional mechanism opportunity, not measured timing or a universal latency improvement. The equal-width initial-state cards center each result identity and caption together with cache-block and lease-lock icons. A closed lock denotes a leased prefix, and the open lock beside C denotes a resident prefix awaiting a lease. The block icons are schematic. The time-scale and admission qualifications are stated in the caption.}
\end{figure*}

\subsection{FIFO prefix leases}
\label{sec:lads-prefix-leases}

Leases apply the admission target $B(t)$ to preserve resident prefixes between tool readiness and reuse. At actual tool-result readiness, the engine-side lease provider updates the target and attempts a lease before exposing the sender descriptor. It considers candidates in ready-time, identity, and version order, keeping protection allocation independent of the sender's priority score. A grant covers the entire reusable prefix returned by the resident-block lookup, with no partial subset chosen for packing.

A grant succeeds only if its $\Delta_i$ additional distinct blocks satisfy $U(t)+\Delta_i\leq B(t)$ and leave at least $H$ free blocks after acquiring the references. Empty resident prefixes receive no lease. The deadline is $d_i=g_i+\ell$. Repeated observations leave an existing lease unchanged. A later regrant cannot reset $g_i$ or extend the lifetime beyond $g_i+\ell$.

The lease provider retries unprotected, still-pending identities from a causal sender inventory in the same FIFO order. The inventory establishes identity and phase; its proposed ranking and time estimates do not select recipients. Replenishment lets returned quota cover another resident history. Inventory delivery can race with commitment, so replenishment records distinguish grants before and after actual commitment. A lower target preserves existing references and accounts for them through $C(t)$ until handoff, expiry, cancellation, or identity invalidation releases them. Target changes neither renew leases nor directly revoke them.

Before allocation, the engine-side lease provider may revoke lease references to preserve headroom. Each revocation removes only that lease's extra references, preserving native references and references held by other leases.

\subsection{Quota turnover priority}

The lease provider publishes lease identities, phases, deadlines, exclusive counts, occupancy, and a delay proxy combining two observed queues. Using the same engine waiting count $w$, let $q_i$ be the zero-based position in the local FIFO queue awaiting engine submission, ordered by arrival and stable identity; for an upstream request absent from that queue, $q_i$ is its current length. This local queue drains each engine iteration through immediate \texttt{add\_request}, without an additional admission limit. Using the same frozen time proxy $\kappa$, the published delay in seconds is
\begin{equation}
 \widehat e_i=\max(1,w+q_i+1)\kappa.
 \label{eq:lads-engine-delay}
\end{equation}
Prior development cold-prefill steps provide $\kappa=50.116$ and $47.155$\,ms on GPU0 and GPU1, respectively, by normalizing elapsed host step time per 512 scheduled tokens. These card-specific values remain fixed across both Web-agent realizations. The proxy uses queue state and batching-inclusive host time without fitting test latency.

At decision time $t$, the sender adds serialization for a feasible frame:
\begin{equation}
 \widehat T_i=\frac{Q+h+p_i}{R}+\widehat e_i,
 \qquad \rho_i=\frac{E_i}{\max(\widehat T_i,\epsilon)},
 \label{eq:lads-score}
\end{equation}
Here $Q$, $h$, and $p_i$ are committed full-frame bytes, framing bytes, and payload bytes; $R$ is bytes/s and $\epsilon=10^{-9}$ seconds guards division. These byte counts use the transmitted representation after any common lossless encoding. Thus $\widehat T_i$ is seconds and $\rho_i$ is blocks/s. Counting a partially transmitted frame at full size conservatively overestimates its remaining serialization. Engine scheduling and future arrivals make total handoff delay an estimate.

The override activates at $C(t)>0$ and $U(t)\geq\theta C(t)$, where $\theta=0.75$ denotes 75.0\% committed protection occupancy, or when fresh evidence shows that a feasible pending result was denied protection because $U(t)+\Delta_i>B(t)$. The latter captures whole-prefix packing pressure below the threshold. Cold prefixes and headroom rejections do not trigger it.

Unlike the FIFO inventory estimate $\widehat T_i^B$, $\widehat T_i$ estimates handoff delay if result $i$ is sent next. Among valid, identity-matching upstream lease owners with $E_i>0$, $0<\widehat e_i<\infty$, and $t+\widehat T_i\leq d_i$, LADS selects the highest $\rho_i$, breaking ties with $k_i$. The score heuristically prioritizes quota turnover; workflow benefit depends on other requests' waiting and cache reuse (Figure~\ref{fig:turnover-example}).

\subsection{Bounded intervention and handoff}

A promotion is a committed departure from Cache-first. The override balances quota turnover with the immediate reuse and short transfers favored by the base order. At most one actual departure from Cache-first is permitted consecutively. The following commitment uses Cache-first, after which the override becomes eligible again. A preview that is not committed changes neither this counter nor byte accounting.

\begin{algorithm}[t]
\caption{\sys{} at a full-frame commitment opportunity}
\label{alg:lads}
\begin{algorithmic}[1]
\Require Feasible set $\mathcal F$, ownership record
\Statex \hspace{\algorithmicindent} Prior-promotion flag $z$
\If{$\mathcal F=\varnothing$}
 \State \Return without commitment
\EndIf
\State $b\gets\arg\min_{i\in\mathcal F}k_i$; $j\gets b$
\If{$z=0$ and record is valid and protection is pressured}
  \State $\mathcal E\gets\{i\in\mathcal F:\text{valid upstream lease},\ E_i>0,$
  \Statex \hspace{\algorithmicindent} $0<\widehat e_i<\infty,\ t+\widehat T_i\leq d_i\}$
 \If{$\mathcal E\neq\varnothing$}
  \State $j\gets\arg\max_{i\in\mathcal E}\rho_i$, ties by $k_i$
 \EndIf
\EndIf
\State Commit the complete frame of $j$ in transport order
\State $z\gets\mathbf 1[j\neq b]$
\end{algorithmic}
\end{algorithm}

Complete receipt binds the observation to its workflow, turn, prefix digest, and payload digest, then immediately registers its request. Receipt and registration do not release protection. The handoff callback runs only after allocation succeeds with positive new-token work and verifies that the native request owns every protected block. It then removes exactly the lease's extra references. Replenishment can use the resulting logical quota while the resumed model call continues running.

\section{Implementation}
\label{sec:implementation}

We implement LADS in Python as a tool-result gateway coupled to a vLLM inference service. In the default evaluations, initial model inputs enter the engine directly, while subsequent tool observations become eligible for delivery when tool execution completes. The gateway chooses among ready observations and transmits complete result frames over a userspace-paced localhost TCP connection. After complete receipt, the continuation enters the engine's FIFO registration path.

Two workload adapters realize this shared Web-agent execution model. The controlled adapter uses archived tool text and resized histories to construct repeatable multi-turn workloads with configurable return cohorts and context demand; Web-derived burst replay feeds prepared Web observations into this adapter. The Web-agent adapter uses a custom ReAct action loop over a local, read-only product and review catalog. Capture records model-selected task paths; timing replay preserves their model inputs, observations, dependencies, and tool-service durations across delivery policies.

The sender maintains the uncommitted inventory and fixed committed-frame order. vLLM tracks per-block reference counts; allocation acquires references to reusable prefix blocks and keeps them out of the free queue~\cite{vllm2025prefix}. LADS performs cache lookup, extra-reference acquisition, expiry, revocation, and handoff on the engine thread, aligning lease changes with allocation. These operations use native block references rather than extra model calls, cache reconstruction, or state transfer.

The engine publishes timestamped, version-bound lease records; the sender uses the latest valid record for each budget generation and falls back to Cache-first on invalid evidence. Appendix~\ref{app:measurement} specifies freshness, identity, and budget-denial checks.

A decision scans candidates and lease records; lease acquisition and release update the extra reference for each protected block.

\section{Evaluation}
\label{sec:evaluation}

We compare complete-workflow latency in controlled bursts, Web-derived burst replay, and captured Web-agent paths, then examine matched component controls, stage timing, and control costs.

\subsection{Experimental setup}
\label{sec:eval-design}
\label{sec:eval-metrics}

\paragraph{Common platform.} All workloads use Qwen2.5-7B-Instruct~\cite{yang2024qwen25} through vLLM on a host with two NVIDIA RTX PRO 5000 Blackwell GPUs, with at most eight concurrent sequences and a 512-token per-step batching budget. The sender uses a four-frame in-flight window and 1028 application-framing bytes per payload: a 1024-byte fixed metadata header plus a 4-byte length prefix. Initial model inputs enter the engine directly; subsequent completed tool results enter delivery scheduling. Initial inference and all tool waits remain in complete-workflow time. Within each timing comparison, methods use the same model, prefix-cache and batching settings, input realization, and physical card. Appendix~\ref{app:measurement} lists exact software versions and hardware details.

\paragraph{Controlled Web-agent realization.} This realization models concurrent Web-agent workflows with configurable histories and tool-return patterns. The default setting contains $N=16$ workflows with six model calls each: 16 cold calls and 80 continuations. Inert archived tool text and resized histories create controlled context competition. Histories have 8,192, 12,288, 20,480, or 22,528 tokens; the context limit is 32,768 tokens, the output cap is 64 tokens, and the cache provides 197,040 usable tokens. Tool service is grouped into two cohorts of $C=8$ workflows. Same-round calls in a cohort start before its common 50-millisecond service completes, with a 60-second limit permitting partial release. Completed results contend for a 32\,kbit/s userspace-paced localhost return link. The default comparison uses five input seeds.

\paragraph{Web-derived burst replay.} Both Web replay workloads use a local, read-only catalog of product and review data from the WebArena shopping database~\cite{zhou2024webarena}, a 32,768-token context limit, and 12,316 configured KV blocks. For this workload, observations from 16 shopping tasks populate the controlled cohort-16 workload shape. Each run retains the controlled realization's 16-workflow, six-call structure (16 initial calls and 80 tool-result continuations), four history lengths, 64-token output cap, and 32\,kbit/s return link. Three input seeds vary workflow arrivals and history-length assignments while preserving each seed's inputs across methods.

\paragraph{Captured Web-agent workflow replay.} This validation records model-selected paths for 16 shopping tasks from the same catalog. Each task asks the agent to inspect three products and their review pages using \texttt{open} and \texttt{read\_reviews} over the local catalog. Histories accumulate from actual observations and model outputs. Replay preserves each path's complete model inputs, observations, dependencies, and independent tool-service durations across methods. The recorded paths contain 204 model calls: 16 initial calls and 188 tool-result continuations. The output cap is 256 tokens with natural stopping. Both methods apply gzip6 lossless compression to tool observations over the same 48\,kbit/s return link, with encoding and decoding included in measured workflow time. The comparison uses three matched arrival seeds over the same captured task paths.

\paragraph{Policies and matched controls.} All policies select completed results at the same gateway. References are ReAct with ready-order FIFO delivery, Size-first (smallest feasible frame), and our delivery-side adaptation of Tail-Risk-Aware Turn Release Scheduling (Tail-aware Release)~\cite{feng2026release}. The adaptation retains its ready-frozen full-prompt work estimate plus four times causal expected output, engine feedback, and model-completion ledger retirement. LADS combines Cache-first ordering, bounded FIFO leases, and turnover. Component controls compare Cache-first without leases, Turnover-off with quota-turnover priority disabled, and complete LADS. LADS and Turnover-off share inputs, model entry, the lease lifecycle, the demand-derived budget rule, headroom $H=2048$ blocks, and cumulative lease limit $\ell=8$ seconds; execution feedback can produce different realized trajectories.

\paragraph{Metrics and aggregation.} The primary metric is mean complete-workflow latency over all $N$ workflows, measured in seconds from workflow arrival to termination. It includes initial inference, tool-service waiting, delivery, engine waiting, and the final model call. Tables average per-run means and nearest-rank P95 values equally. With 16 workflows, nearest-rank P95 selects the largest observation. Each reported percentage is computed within a matched seed and then averaged equally: $\overline\Delta=M^{-1}\sum_{k=1}^{M}100(T_{b,k}-T_{c,k})/T_{b,k}$, where $M$ is the number of matched seeds. Positive values indicate lower candidate latency. This differs from a ratio of displayed mean times. Makespan measures whole-batch completion separately from per-workflow latency. Captured-workflow statistics average three matched arrival seeds over the same recorded task paths.

Cache measurements distinguish historical-prefix lookup hints from engine full-prompt cache observations. Controller CPU time is cumulative CPU seconds per run, including budget evaluation. Protection block-time is the time integral of the distinct protected-block count $U(t)$, measured in block-seconds. Serialized control bytes, protection block-time, preemptions, grants, expiries, and handoffs accompany latency. Each method's control and protection mechanisms are included in its measured cost. Actual allocation, handoff, and expiry events establish what occurred when published state lags execution.

\begin{table}[!htbp]
\centering\small\setlength{\tabcolsep}{4pt}
\caption{Complete-workflow latency (seconds; lower is better). Controlled cohort-8 results average five input seeds; Web-derived burst replay averages three input seeds. The two workload columns use 16 workflows, 96 calls, and a 32 kbit/s return link. Entries average run-level means and P95s equally. Boldface identifies LADS; underlining marks minima.}
\label{tab:lads-eval-main}
\begin{tabular}{lrrrr}
\toprule
 & \multicolumn{2}{c}{\shortstack{Controlled\\(cohort-8)}} & \multicolumn{2}{c}{\shortstack{Web-derived\\burst replay}}\\
\cmidrule(lr){2-3}\cmidrule(lr){4-5}
Method & Mean & P95 & Mean & P95\\
\midrule
ReAct (FIFO) & 196.6 & 210.4 & 204.8 & 218.1\\
Size-first & 186.7 & 201.1 & 158.1 & 171.6\\
Tail-aware Release & 176.5 & 196.7 & 192.3 & 211.4\\
\textbf{LADS (ours)} & \textbf{\underline{108.9}} & \textbf{\underline{119.2}} & \textbf{\underline{118.0}} & \textbf{\underline{130.0}}\\
\bottomrule
\end{tabular}
\end{table}

\subsection{End-to-end workflow performance}
\label{sec:eval-performance}
\label{sec:eval-controlled}
\label{sec:eval-replay}

Table~\ref{tab:lads-eval-main} shows that LADS lowers mean complete-workflow latency in controlled cohort-8 by 44.6\% against FIFO, 41.7\% against Size-first, and 37.0\% against Tail-aware Release across five matched input seeds. At cohort-16, LADS also achieves the lowest mean latency: 119.0 seconds versus 204.6 seconds for FIFO (Table~\ref{tab:lads-latest-main}).

In Web-derived burst replay, the corresponding reductions are 42.4\%, 25.3\%, and 38.6\% across three matched seeds, with LADS improving on every reference in every seed. These totals include all workflows, initial calls, and tool waits; Table~\ref{tab:lads-eval-main} reports absolute means and P95s.

\label{sec:eval-replication}

\begin{table}[!htbp]
\centering\small
\caption{Complete-workflow latency for captured Web-agent paths (seconds). Both methods replay the same 16 recorded paths, comprising 204 model calls and 188 continuations, using gzip6 at 48 kbit/s. Entries equally average three matched arrival-seed run statistics. Boldface identifies LADS (ours).}
\label{tab:lads-captured-trace-pair}
\begin{tabular}{lrr}
\toprule
Method & Mean & P95\\
\midrule
ReAct (FIFO) & 156.6 & 187.4\\
\textbf{LADS (ours)} & \textbf{77.3} & \textbf{105.1}\\
\bottomrule
\end{tabular}
\end{table}

\subsection{Captured Web-agent workflow replay}
\label{sec:eval-captured-trace}

Table~\ref{tab:lads-captured-trace-pair} evaluates FIFO and LADS on the 16 captured task paths. Across three matched arrival seeds, LADS reduces mean complete-workflow latency by 50.6\% (48.2--51.8\% within seeds), with lower latency in all 48 paired workflow comparisons. Mean latency falls from 156.6 to 77.3 seconds, including initial inference and all tool waits.

\begin{figure}[htbp]
\centering
\includegraphics[width=.95\columnwidth,trim=2bp 0bp 3bp 2bp,clip]{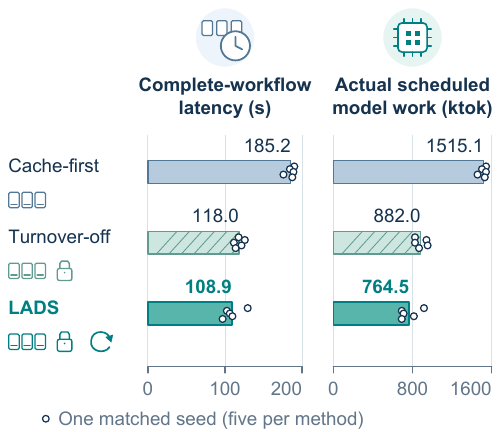}
\caption{Controlled cohort-8 component ablation (16 workflows, five matched input seeds). Bars show equal-weight means; points show individual runs. Cache-first has no leases; Turnover-off shares LADS's budget and lease lifecycle with quota-turnover priority disabled. Model work includes prefill, decode, and repeated execution.}
\label{fig:lads-component-ablation}
\Description{Two side-by-side metric panels compare Cache-first, Turnover-off, and LADS in three rows. The left panel reports mean complete-workflow latency in seconds; the right panel reports actual scheduled model work in thousands of tokens. Each axis starts at zero with an independent scale. Bars show means, with five matched-seed points per method and metric. Cache-block, lock, and turnover pictograms identify cache ordering, prefix leases, and quota-turnover priority.}
\end{figure}

\subsection{Ablation and mechanism analysis}
\label{sec:eval-component}

Figure~\ref{fig:lads-component-ablation} separates protection (Cache-first versus Turnover-off) from turnover (Turnover-off versus complete LADS). With the budget rule and lease lifecycle held constant, turnover lowers mean cohort-8 latency by 7.6\% across five matched inputs, improving four seeds.

\begin{figure}[tbp]
\centering
\includegraphics[width=.95\columnwidth]{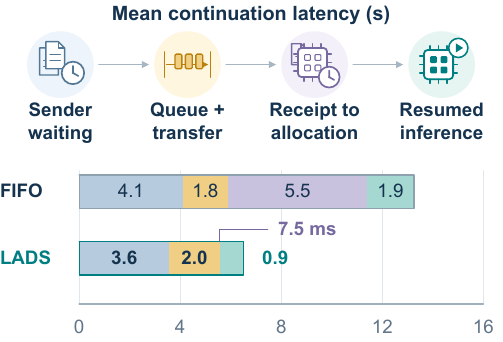}
\caption{Captured Web-agent continuation latency (gzip6, 48\,kbit/s). Stage means use the same 188 continuations per run and weight the three matched arrival seeds equally. Receipt-to-allocation includes registration and engine waiting. Initial calls are included in complete-workflow totals.}
\label{fig:lads-captured-web-latency}
\Description{A unified figure contains four connected schematic pictograms identifying the stages: queued result cards with a clock, packet transfer along a link, an inference engine with a waiting clock, and an active inference engine. FIFO and LADS profiles concatenate the four mean stage durations on a common 0--16 second scale. Colors connect the pictograms to the corresponding measured stages. FIFO durations are 4.1, 1.8, 5.5, and 1.9 seconds; LADS durations are 3.6 seconds, 2.0 seconds, 7.5 milliseconds, and 0.9 seconds. Pictogram positions and sizes are schematic; profile segment widths encode measured stage means.}
\end{figure}

For the captured paths, Figure~\ref{fig:lads-captured-web-latency} shows receipt-to-allocation falling from 5.5 seconds to 7.5 milliseconds per continuation and allocation-to-completion from 1.9 to 0.9 seconds. Engine full-prompt cache observations rise from 0.95M to 1.69M cached tokens per run, alongside 51.7 promotions and 95.3 verified handoffs. Tables~\ref{tab:lads-captured-web-local} and~\ref{tab:lads-captured-web-cost} give all intervals and costs from these runs.

\subsection{Control overhead}
\label{sec:eval-cost}
\label{sec:eval-closed-loop}

In captured Web-agent replay, LADS uses 11.5 cumulative controller CPU seconds per run versus FIFO's 8.4, including budget evaluation and state synchronization. It consumes 144,792.6 protection block-seconds per run; all 148,381 added references across the three runs are released. Codec time remains included. Tables~\ref{tab:lads-nonweb-ablation} and~\ref{tab:lads-captured-web-cost} report the remaining costs; serialized control bytes measure in-process records.

\section{Related Work}
\label{sec:related}

Web agents use returned observations to guide subsequent model decisions~\cite{yao2023react,zhou2024webarena}. Serving such agents concurrently involves scheduling model calls and managing reusable context~\cite{luo2026agentix,abhyankar2024infercept}.

\paragraph{Tool use and Web-agent evaluation.} ReAct interleaves reasoning and action~\cite{yao2023react}; Toolformer, ToolLLM, and Gorilla develop tool/API use~\cite{schick2023toolformer,qin2024toolllm,patil2024gorilla}, while LLMCompiler exposes parallel-call dependencies~\cite{kim2024llmcompiler}. Interactive and action-generation benchmarks include WebShop, WebArena, Mind2Web, WebLINX, VisualWebArena, and WorkArena~\cite{yao2022webshop,zhou2024webarena,deng2023mind2web,lu2024weblinx,koh2024visualwebarena,drouin2024workarena}; BrowserGym, AgentLab, and AgentBoard provide evaluation and progress measurement~\cite{chezelles2025browsergym,servicenow2026agentlab,ma2024agentboard}. LADS schedules completed observations across workflows while holding agent policy and observation semantics common in matched comparisons.

\paragraph{Inference scheduling and application structure.} Orca, Sarathi-Serve, and DistServe coordinate iterations, prefills, and decoding~\cite{yu2022orca,agrawal2024sarathi,zhong2024distserve}; FastServe and Llumnix preempt or redistribute requests~\cite{wu2026fastserve,sun2024llumnix}. AlpaServe and the Virtual Token Counter address allocation and fairness~\cite{li2023alpaserve,sheng2024vtc}, while PagedAttention and vAttention manage physical cache memory~\cite{kwon2023pagedattention,prabhu2025vattention}. SGLang, Parrot, and Agentix exploit program structure or cross-call progress~\cite{zheng2024sglang,lin2024parrot,luo2026agentix}; Sutradhara co-designs tool orchestration and inference to overlap tools with prefill~\cite{biswas2026sutradhara}, and Preble combines prefix sharing with placement~\cite{srivatsa2025preble}. LADS instead selects which completed result reaches ordinary FIFO registration next.

\paragraph{Context reuse and residency.} Prompt Cache, CachedAttention, and ChunkAttention reuse modular, multi-turn, or shared-prefix state~\cite{gim2024promptcache,gao2024cachedattention,ye2024chunkattention}; CacheBlend combines cached knowledge segments~\cite{yao2025cacheblend}. CacheGen and Mooncake compress, stream, or pool KV state~\cite{liu2024cachegen,qin2025mooncake}, and Pensieve spans memory tiers~\cite{yu2025pensieve}. PEEK guides admission and eviction with pending prefixes~\cite{xie2026peek}; H$_2$O, StreamingLLM, and KIVI select or quantize retained context~\cite{zhang2023h2o,xiao2024streamingllm,liu2024kivi}. LADS couples completed-result order with bounded resident-prefix protection and quota return at handoff.

\paragraph{Interruptions, readiness, and delivery.} InferCept manages context around external calls, Continuum uses cache time-to-live, and Ask the Tool, Don't Guess uses tool progress for retention/reload decisions~\cite{abhyankar2024infercept,li2026continuum,liu2026asktool}. Tail-aware Release prioritizes ready turns and controls committed model work~\cite{feng2026release}; our delivery-side adaptation retains its work estimate, causal output history, engine feedback, and model-completion ledger. SCOOT tunes inference-engine configurations~\cite{cheng2025scoot}, Baton reorganizes batches~\cite{cong2025baton}, and Fate accelerates edge mixture-of-experts inference~\cite{fang2026fate}. LADS addresses completed-observation delivery and reuse at handoff.

\section{Discussion}
\label{sec:discussion}

LADS targets shared return paths where sender backlog coincides with resident histories competing for finite protection. The benefit depends on whether an owner reaches handoff within its lease horizon; sustained committed transfers and engine waiting can consume that horizon before useful turnover.

The evaluation covers randomized controlled inputs, Web-derived burst replay, and captured model-selected paths using one model. Web-derived input seeds vary arrivals and history assignments, while captured-path arrival seeds replay the same task paths. Broader tasks and models would test how return-link capacity, reuse, and lease duration shape this operating regime.

\section{Conclusion}
\label{sec:conclusion}

LADS is a tool-result delivery scheduler for concurrent Web agents sharing a constrained return path and an inference service. It protects waiting histories with bounded prefix leases and, under protection pressure, prioritizes results whose timely resumption can return protection quota to other waiting histories. Compared with ReAct using FIFO tool-result delivery, LADS reduces mean complete-workflow latency by 44.6\% on controlled cohort-8 bursts and 42.4\% on Web-derived burst replay, without changing the model or agent policy. In the two burst workloads, LADS lowers mean latency relative to all three delivery baselines on every input seed.

More broadly, LADS makes tool-result delivery part of agent serving: delivery order affects when requests reach inference and whether cached histories remain available for reuse. As Web agents undertake longer tasks with more tool calls on shared inference services, coordinating communication with inference state offers a practical way to improve serving efficiency.

\flushcolsend
\clearpage
\bibliographystyle{ACM-Reference-Format}
\bibliography{references}
\appendix
\section{Additional evaluation results}
\label{app:exact-main-components}

\begin{table}[!htbp]
\centering\small\setlength{\tabcolsep}{4pt}
\caption{Five-seed controlled complete-method comparisons, in complete-workflow seconds. A/B denote cohort-8/cohort-16, with 16 workflows per run. Entries equally average five run-level means or nearest-rank P95 values. Boldface identifies LADS (ours).}
\label{tab:lads-latest-main}
\begin{tabular}{lrrrr}
\toprule
Method & A mean & A P95 & B mean & B P95\\
\midrule
ReAct (FIFO) & 196.6 & 210.4 & 204.6 & 218.3\\
Size-first & 186.7 & 201.1 & 151.6 & 165.1\\
Tail-aware Release & 176.5 & 196.7 & 203.4 & 223.1\\
\textbf{LADS (ours)} & \textbf{108.9} & \textbf{119.2} & \textbf{119.0} & \textbf{131.7}\\
\bottomrule
\end{tabular}
\end{table}

\begin{table*}[!htbp]
\centering\small\setlength{\tabcolsep}{4pt}
\caption{Controlled cohort-8 component ablation over five matched input seeds. Work is actual scheduled model work, including prefill, decode and repeat work, in thousands of tokens; protection is thousands of distinct-block seconds. Actions are promotions/handoffs/expiries, averaged per run. CPU includes the complete controller. Cache-first has no added leases; Turnover-off and LADS share the demand-budget rule and lease lifecycle, differing only in sender turnover. Boldface identifies LADS (ours).}
\label{tab:lads-nonweb-ablation}
\begin{tabular}{lrrrrr}
\toprule
Method & Mean (s) & Work (ktok) & Protect. (kblock-s) & Actions & CPU (s)\\
\midrule
Cache-first & 185.2 & 1515.1 & 0 & 0/0/0 & 1.7\\
Turnover-off & 118.0 & 882.0 & 271.3 & 0/28.6/17.8 & 3.5\\
\textbf{LADS (ours)} & \textbf{108.9} & \textbf{764.5} & \textbf{274.0} & \textbf{21.6/36.4/18.2} & \textbf{4.3}\\
\bottomrule
\end{tabular}
\end{table*}

\begin{table*}[!htbp]
\centering\small\setlength{\tabcolsep}{3.5pt}
\caption{Mean stage latency per continuation in captured Web-agent workflow replay. Units are shown in the column headers. Entries equally average three matched run means over the same 188 continuations per run. Receipt-to-allocation includes registration and engine waiting. Boldface identifies LADS (ours).}
\label{tab:lads-captured-web-local}
\begin{tabular}{lrrrr}
\toprule
Method & \shortstack{Ready--\\commit (s)} & \shortstack{Commit--\\receipt (s)} & \shortstack{Receipt--\\alloc. (ms)} & \shortstack{Alloc.--\\end (s)}\\
\midrule
ReAct (FIFO) & 4.1 & 1.8 & 5500.3 & 1.9\\
\textbf{LADS (ours)} & \textbf{3.6} & \textbf{2.0} & \textbf{7.5} & \textbf{0.9}\\
\bottomrule
\end{tabular}
\end{table*}

\begin{table*}[!htbp]
\centering\small\setlength{\tabcolsep}{4pt}
\caption{Cache reuse and control costs in captured Web-agent workflow replay. Entries equally average three per-run measurements. Cache input covers all 204 calls; missing input is prompt tokens minus cached tokens. CPU includes state-synchronization checks and owner budget control. P/H/E gives mean per-run promotions, verified handoffs, and expiries. Boldface identifies LADS (ours).}
\label{tab:lads-captured-web-cost}
\begin{tabular}{lrrrrrrr}
\toprule
Method & Cached (ktok) & Missing (ktok) & Protect. (kblock-s) & CPU (s) & Budget CPU (s) & Control (MiB) & P/H/E\\
\midrule
ReAct (FIFO) & 947.9 & 1053.9 & 0 & 8.4 & 0 & 11.6 & 0/0/0\\
\textbf{LADS (ours)} & \textbf{1693.2} & \textbf{308.6} & \textbf{144.8} & \textbf{11.5} & \textbf{5.3} & \textbf{9.8} & \textbf{51.7/95.3/9}\\
\bottomrule
\end{tabular}
\par\smallskip{\raggedright\footnotesize LADS budget-context observation uses 0.3 CPU seconds per run, included in total CPU. FIFO/LADS codec CPU is 40.8/40.7 ms per run, recorded separately and included in workflow wall time.\par}
\end{table*}

\begin{table*}[!p]
\centering\small\setlength{\tabcolsep}{4pt}
\caption{Per-seed complete-workflow means (seconds). Controlled cohort-8 and cohort-16 comparisons each use five input seeds; Web-derived burst replay uses three input seeds. Each group preserves its own seed identity. Boldface identifies LADS (ours).}
\label{tab:lads-latest-main-seeds}
\begin{tabular}{llrrrr}
\toprule
Workload & Seed & FIFO & Size-first & Tail-aware Release & \textbf{LADS (ours)}\\
\midrule
Cohort-8 & 1667935512 & 194.9 & 187.5 & 122.4 & \textbf{96.9}\\
Cohort-8 & 1461454527 & 194.4 & 175.8 & 182.6 & \textbf{109.6}\\
Cohort-8 & 1055731457 & 196.6 & 184.4 & 193.0 & \textbf{102.4}\\
Cohort-8 & 1321697771 & 197.9 & 189.1 & 196.3 & \textbf{106.2}\\
Cohort-8 & 711104771 & 199.4 & 196.6 & 188.1 & \textbf{129.5}\\
Cohort-16 & 1667935512 & 201.0 & 157.8 & 197.1 & \textbf{111.6}\\
Cohort-16 & 1461454527 & 200.2 & 143.8 & 190.8 & \textbf{110.5}\\
Cohort-16 & 1055731457 & 210.9 & 149.8 & 209.5 & \textbf{116.1}\\
Cohort-16 & 1321697771 & 203.3 & 152.1 & 201.2 & \textbf{119.6}\\
Cohort-16 & 711104771 & 207.7 & 154.4 & 218.3 & \textbf{137.3}\\
Web-derived burst & 732418792 & 200.2 & 159.0 & 196.7 & \textbf{112.4}\\
Web-derived burst & 1533424038 & 207.4 & 159.4 & 188.4 & \textbf{125.0}\\
Web-derived burst & 55546983 & 206.9 & 155.7 & 191.7 & \textbf{116.7}\\
\bottomrule
\end{tabular}
\end{table*}

\paragraph{Captured-path stage measurements.}
\label{app:web-details}

Tables~\ref{tab:lads-captured-web-local} and~\ref{tab:lads-captured-web-cost} report stage timing, cache work, and control costs for the same three matched arrival seeds in the captured Web-agent workflow replay of Section~\ref{sec:eval-captured-trace}.

\paragraph{Randomized inputs and replication pairs.}
\label{app:replication}

Table~\ref{tab:lads-latest-main-seeds} lists the five input seeds for each controlled cohort setting and the three Web-derived burst input seeds. Table~\ref{tab:lads-captured-trace-seeds} lists the three matched arrival seeds over the same captured task paths. Each group preserves its own seed identity.

\section{Budget estimation and admission}
\label{app:budget}

\label{app:budget-estimator}

Section~\ref{sec:lads-budget-estimation} defines the budget estimate and near-use demand. The lease provider uses its current local inventory or the latest accepted, identity-matching sender inventory. The FIFO prefix contains the continuation frames currently carried by the return path. A partly transmitted frame can conservatively remain counted at full size.

\section{Measurement and validity checks}
\label{app:measurement}

Table~\ref{tab:reproducibility} lists the software versions and hardware configuration shared by all workloads.

\begin{table*}[t]
\begin{minipage}[t]{\columnwidth}
\centering\small\setlength{\tabcolsep}{4pt}
\caption{Per-arrival-seed complete-workflow means for captured Web-agent workflow replay (seconds). FIFO and LADS use the same recorded task paths and matched arrivals across three seeds. Boldface identifies LADS (ours).}
\label{tab:lads-captured-trace-seeds}
\begin{tabular}{rrr}
\toprule
Arrival seed & FIFO & \textbf{LADS (ours)}\\
\midrule
172910251 & 159.0 & \textbf{76.7}\\
1147784114 & 159.7 & \textbf{76.9}\\
688237994 & 151.1 & \textbf{78.2}\\
\bottomrule
\end{tabular}

\end{minipage}\hfill
\begin{minipage}[t]{\columnwidth}
\caption{Software and hardware configuration shared by all workloads.}
\label{tab:reproducibility}
\centering
\small
\begin{tabular}{@{}p{0.32\columnwidth}p{0.62\columnwidth}@{}}
\toprule
Component & Configuration \\
\midrule
Model revision & \texttt{a09a354} \\
PyTorch & 2.8.0 \\
vLLM & 0.11.0 \\
Transformers & 4.56.2 \\
Host CPU & AMD Ryzen Threadripper PRO 7975WX \\
GPU model & NVIDIA RTX PRO 5000 Blackwell \\
Reported GPU memory & 48,935\,MiB per card \\
\bottomrule
\end{tabular}

\end{minipage}
\end{table*}

\paragraph{Stage latencies.} Using the timestamps in Section~\ref{sec:return-path}, we divide each continuation's ready-to-completion time into sender waiting ($s-r$), queuing and transfer ($c-s$), receipt to allocation ($n-c$, including registration and engine waiting), and resumed inference ($f-n$). Table~\ref{tab:lads-captured-web-local} and Figure~\ref{fig:lads-captured-web-latency} average the same continuations within each run, then weight run means equally. Initial calls enter complete-workflow latency but not these stage means.

\paragraph{Record validity.} The sender accepts ownership records only when workflow/turn versions and prefix/payload digests match and the records are at most one second old. Target, commitment, and budget-denial records must share the same received budget generation. The sender uses Cache-first when identity, sequence, or timestamp checks fail, including stale or duplicate records, future timestamps, and inconsistent event order. Each completed run verifies that all added lease references have been released.

\flushcolsend
\end{document}